\documentclass[fleqn,10pt,twocolumn]{wlscirep}

\usepackage[utf8]{inputenc}
\usepackage[T1]{fontenc}
\usepackage{bm}
\usepackage{changepage}
\usepackage{adjustbox}
\usepackage{pifont}
\newcommand{\cmark}{\ding{51}}%
\newcommand{\xmark}{\ding{55}}%
\usepackage{multirow}
\usepackage{geometry}
 \definecolor{darkgreen}{rgb}{0.0, 0.5, 0.15}
\usepackage{lipsum}
\usepackage{graphicx}
\RequirePackage{amsmath,amssymb}

\usepackage{verbatim}    % Comment environment
\usepackage{pgf}         % Package defining color names (and some other stuff I don't use)
\usepackage{subcaption}   % Side-by-side figures
\usepackage{placeins}    % FloatBarrier

\usepackage{tcolorbox} %box with text

\newcommand\wilkscond[1]{\textsc{#1}}
\newcommand\wc[1]{\wilkscond{#1}} %:)

\usepackage{amsthm}        % Theorems without numbers
\newtheorem*{theorem}{Theorem}

\title{Searching for new physics with profile likelihoods: Wilks and beyond}

\author[1]{Sara Algeri}
\author[2]{Jelle Aalbers}
\author[2,3]{Knut Dundas Mor{\aa}}
\author[2,*]{Jan Conrad}
\affil[1]{School of Statistics, University of Minnesota, Minneapolis (MN), 55455, USA}
\affil[2]{Physics Department and Oskar Klein Centre, Stockholm University, Stockholm,  Sweden}
\affil[3]{Physics Department, Columbia University,
New York, NY 10027, USA}

\affil[*]{e-mail: conrad@fysik.su.se}

\begin{abstract}
Particle physics experiments use likelihood ratio tests extensively to compare hypotheses and to construct confidence intervals. Often, the null distribution of the likelihood ratio test statistic is approximated by a $\chi^2$ distribution, following a theorem due to Wilks. However, many circumstances relevant to modern experiments can cause this theorem to fail. In this paper, we review how to identify these situations and construct valid inference.
\end{abstract}

\begin{document}

\flushbottom
\maketitle

\thispagestyle{empty}

\begin{tcolorbox}[title=Key points, title filled=false,boxrule=0mm]
Necessary conditions for the Likelihood Ratio Test to be approximated  $\chi^2$ include the following.
\begin{itemize}
    \item The sample size must be sufficiently large. If not, higher-order asymptotic results exist or one can rely on Monte Carlo simulations.
    \item The true values of the parameters must lie in the interior of the parameter space. If not, both asymptotic results and simulations must be adjusted accordingly.
    \item The parameters must be identifiable. If not, look-elsewhere effect correction methods can be of help to address this problem.
    \item The models under comparison must be nested. If not, one can construct a comprehensive model which includes the models considered as special cases.
    \item The models under comparison must be specified correctly. If not, adding nuisance parameters can be of help, otherwise one must rely on nonparametric methods.
\end{itemize}
\end{tcolorbox}

\noindent \textbf{Website summary:} The goal of this manuscript is to identify situations in particle physics
where the likelihood ratio test statistic cannot approximated by a $\chi^2$ distribution and  propose adequate solutions.
%flushbottom  % ??
\maketitle
\thispagestyle{empty}

\section{Introduction}

Modern particle physics employs elaborate statistical methods to enhance the physics reach of expensive experiments and large collaborations. The field’s standard for reporting results is frequentist hypothesis testing. At its heart, this is a two-step recipe for distinguishing a null hypothesis $H_0$ (e.g.~the standard model) from a more general case $H_1$ (e.g.~the standard model with a new particle):

\begin{tcolorbox}[title filled=false, boxrule=0mm, title=Frequentist hypothesis testing]
    \begin{enumerate}
    \item Summarize the data collected by the experiment  with a test statistic $T$ whose absolute value is smaller under $H_0$ than it is under $H_1$.
    \item Compute a p-value, i.e., the probability that $T$ is larger than its observed value when $H_0$ is true. $H_0$ is rejected if the \emph{p-value} is below some threshold $\alpha$ (e.g. $2.9 \times 10^{-7}$, corresponding to a $5 \sigma$ significance level).
    \end{enumerate}
\end{tcolorbox}

Commonly, $H_0$ differs from $H_1$ for some fixed values of the \emph{parameter(s) of interest} which we denote by $\mu$, i.e.,
\begin{equation}
\label{test}
H_0 : \mu = \mu_0 \quad\text{versus}\quad H_1: \mu > \mu_0 \quad \text{(or $\mu\neq \mu_0$)}.
\end{equation}

The model under study may also be characterized by a set of  \emph{nuisance parameter(s)}, namely $\theta$, whose value is not of direct interest for the test being conducted, but it still needs to estimated under both hypotheses. 
For example, in a search for a new physics, $\mu$ may correspond to the rate or cross-section of signal events and $\theta$ could include detector efficiencies or uncertain background rates.  To assess the presence of the signal  we test

The first step of hypothesis testing requires the specification of a test statistic $T$. Among the wide variety of testing procedures available in statistical literature, a popular choice is the profile (or generalized) \emph{Likelihood Ratio Test} (LRT), due to its statistical power \cite{NP,KR} and simple implementation. The LRT has been used in several seminal studies in particle physics, from the discovery of the Higgs boson \cite{atlashiggs,cmshiggs} and measurements of the neutrino properties \cite{dayabay} to direct \cite{xenon1t,pandax,lux} and indirect \cite{hessline,fermilines} searches for dark matter.

Given a \emph{likelihood function}, $L$, that measures the probability of the observed data for a given value of the parameters $\mu$ and $\theta$, the LRT is defined as
\begin{equation}
    \label{LRT}
    T = -2 \log \frac{
		L( \mu_0, \widehat{\theta}_0 )	   }{
		L( \widehat{\mu}, \widehat{\theta})           }
	.
\end{equation}
Here $\mu_0$ is the value of $\mu$ specified under the null hypothesis in \eqref{test}, whereas $\widehat{\theta}_0$ is the best-fit of the nuisance parameter under $H_0$, i.e., the value of $\theta$ that maximize $L$ given that $\mu=\mu_0$. Similarly, $\widehat{\mu}$ and $\widehat{\theta}$ are the best-fit parameters under $H_1$. In the new physics signal search example, $\mu_0 = 0$ (i.e., the only  value of $\mu$ allowed under the null hypothesis zero); it follows that $T$ is zero if $\widehat{\mu} = 0$, and it is positive otherwise.

The second step of hypothesis testing 
is to quantify the evidence in favor of $H_1$ by means of a p-value, i.e., the probability that, when the null hypothesis is true, $T$ is much larger than its value observed on the data. Finally, $H_0$ is excluded if the p-value if smaller than the predetermined probability of false discovery $\alpha$. 

An equivalent statement on the validity of $H_0$ can be made  
by constructing a confidence interval for $\mu$.
The latter corresponds to the set of possible values $\mu_0$ for which the probability that
$T$ is smaller or equal than its value observed is $1-\alpha$. An experiment can exclude a certain value $\mu_0$ at a confidence level (or \emph{coverage}) $1 - \alpha$ if such value is not contained in the confidence interval.

Both test of hypothesis and confidence intervals rely on the distribution of $T$ under $H_0$. A famous result from Wilks \cite{wilks} states that the null  distribution of the LRT in \eqref{LRT} is approximately $\chi^2$ distributed if sufficient data is acquired, provided some regularity conditions are met.
These same results provide the underpinning for the common $\chi^2$ fitting and goodness-of-fit computations. Therefore, they are subject to the same regularity conditions, which, if incorrectly assumed, can yield to invalid claims of exclusions or discoveries.

%While many physicists perform ‘chi-squared’ fits, few realize these are applications of the LRT with Wilks' theorem. Fewer still can state the required regularity conditions, or intuit how much data is needed until Wilks’ result applies. However, when Wilks’ theorem is incorrectly assumed, invalid claims of exclusion and discovery can result.

In this paper, we present a set of necessary conditions for Wilks’ theorem to hold using language and  examples familiar to particle physicists. Furthermore, when they fail, we give recommendations on when extensions of Wilks’ result apply. 

Generally speaking, a viable alternative when Wilks or similar results fail to hold is to estimate the null distribution of $T$  by \emph{Monte Carlo} or \emph{toy} simulations, i.e. simulating many datasets under $H_0$, and computing $T$ for each. However, as discussed in Sections \ref{violation2} and \ref{recommend}, besides the computational burden, additional complications may arise and consistecy of the numerical solution is not guaranteed.

\section{Wilks' theorem and conditions}

Wilks' theorem \cite{wilks} can be stated as follows:

\begin{theorem}[Wilks, 1938]
\label{theo}
Under suitable regularity conditions, when $H_0$ in \eqref{test} is true, the distribution of $T$ converges to $\chi^2_m$. 
\end{theorem}   

Here $\chi^2_m$ is a chi-squared random variable with degrees of freedom $m$ equal to the number of parameters in $\mu$. Instructions on constructing confidence regions and tests based on Wilks' theorem can be found elsewhere \cite{cowan}; here, we focus on the conditions needed for this theorem to apply. A formal statement of the regularity conditions required by Wilks can be found in \cite{coxhinkley} and \cite{protassov}, but five necessary conditions cover most practical cases:

\begin{tcolorbox}[title filled=false, boxrule=0mm, title=Necessary conditions for Wilks' theorem]
    \begin{itemize}
    \item[] \wilkscond{Asymptotic}: Sufficient data is observed.
    \item[] \wilkscond{Interior}: Only values of $\mu$ and $\theta$ which are far from the boundaries of their parameter space are admitted.
    \item[] \wilkscond{Identifiable}: Different values of the parameters specify distinct models.
    \item[] \wilkscond{Nested}: $H_0$ is a limiting case of $H_1$, e.g.~with some parameter fixed to a sub-range of the entire parameter space.
    \item[] \wilkscond{Correct}: The true model is specified either under $H_0$ or under $H_1$.
    \end{itemize} 
\end{tcolorbox}

\begin{figure}
    \centering
    \includegraphics[width=0.5\textwidth]{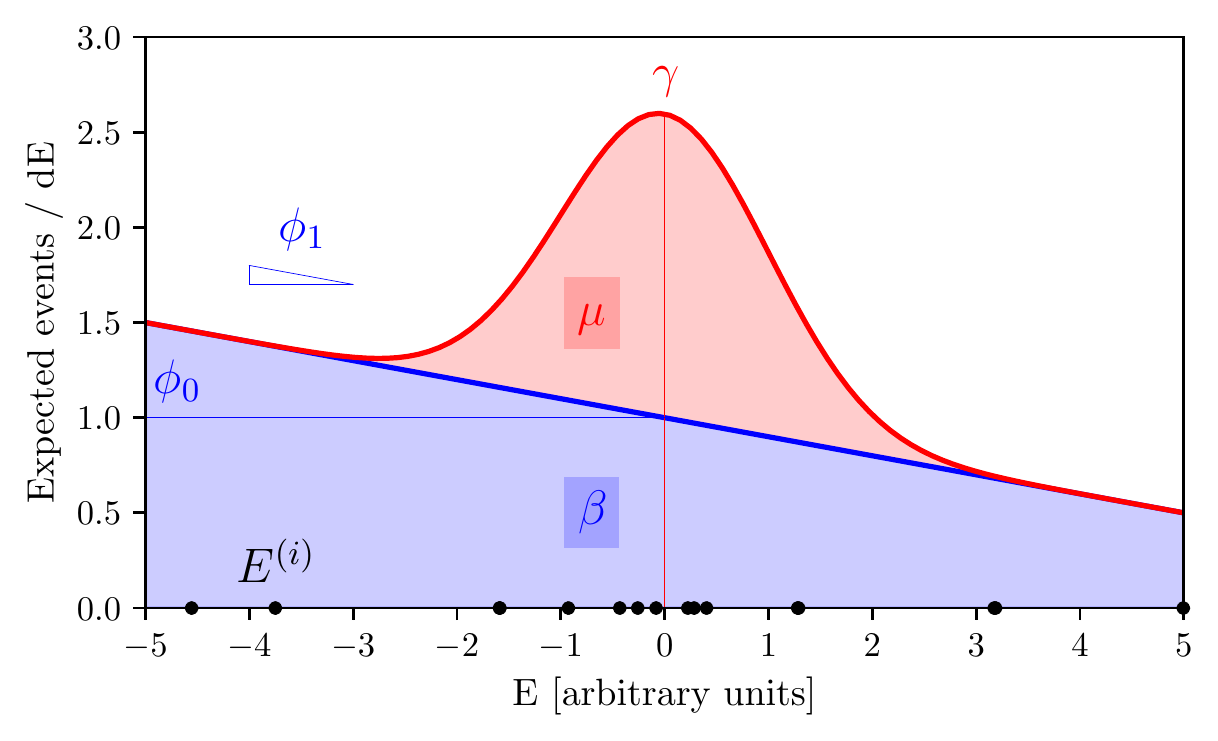}
    \caption{An illustration of the example model used in this paper. The figure shows the distributions of energies of particles measured by an experiment. In the analysis, the aim is to constrain the mean number $\mu$ of signal events from the Gaussian signal (red) centered at $E=\gamma$, on top of the polynomial background (blue) with distribution $B(E) = \phi_0 + \phi_1 E + \ldots$ and $\beta$ expected number of events. The model shown here is $\mu = 4$, $\beta = 10$, $\gamma=0$, $\phi_0=1$, $\phi_1=-0.1$, $\{\phi_j\}_{j\geq2}=0$. The black dots show one dataset of events $\{E^{(i)}\}_{i=1}^N$ that the experiment might observe under this model.}
    \label{model_sketch}
\end{figure}

\begin{figure*}[t]
\centering
\begin{subfigure}[t]{0.5\textwidth}
    \centering
    \includegraphics[width=\textwidth]{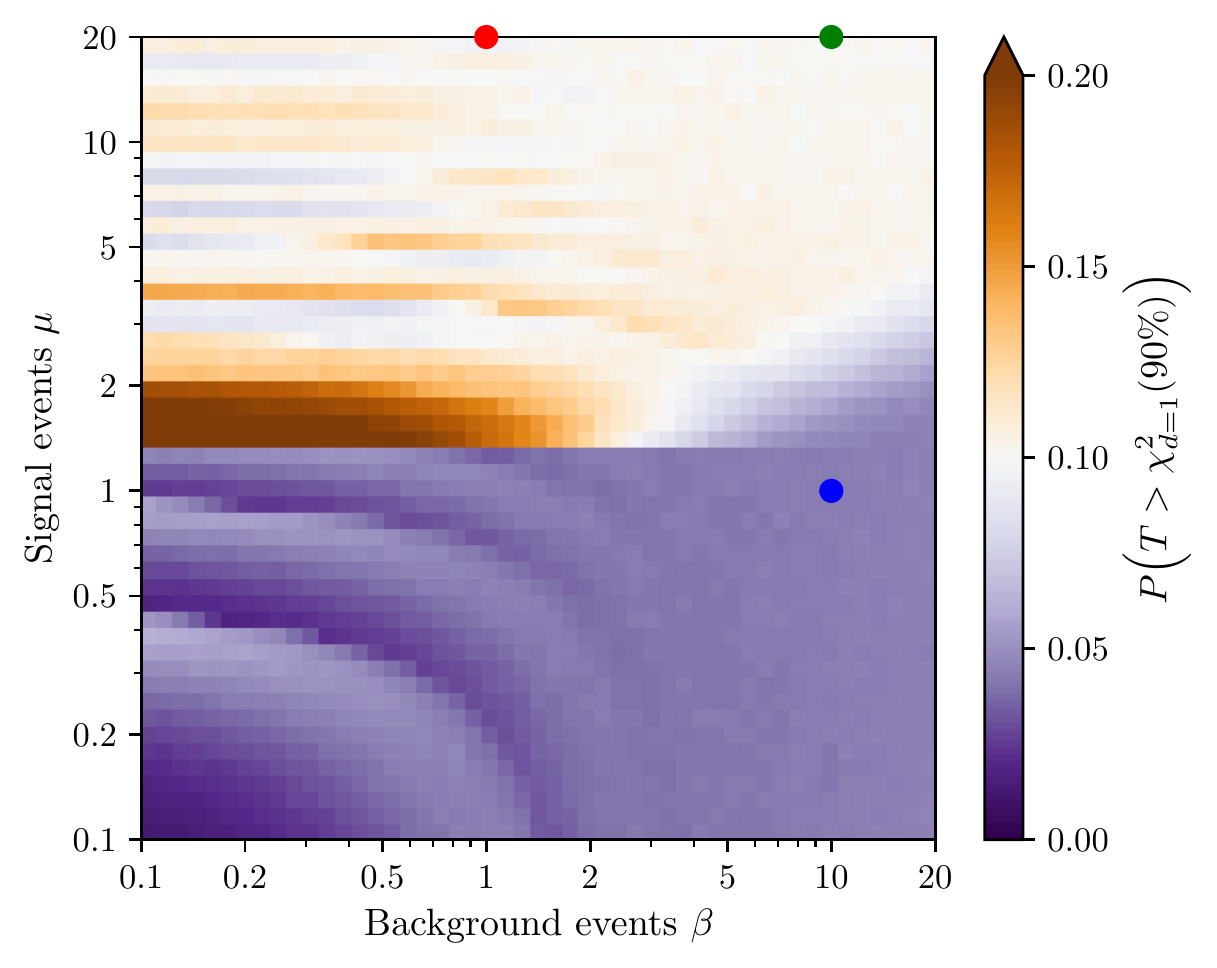}
\end{subfigure}%
\begin{subfigure}[t]{0.5\textwidth}
    \centering
    \includegraphics[width=\textwidth]{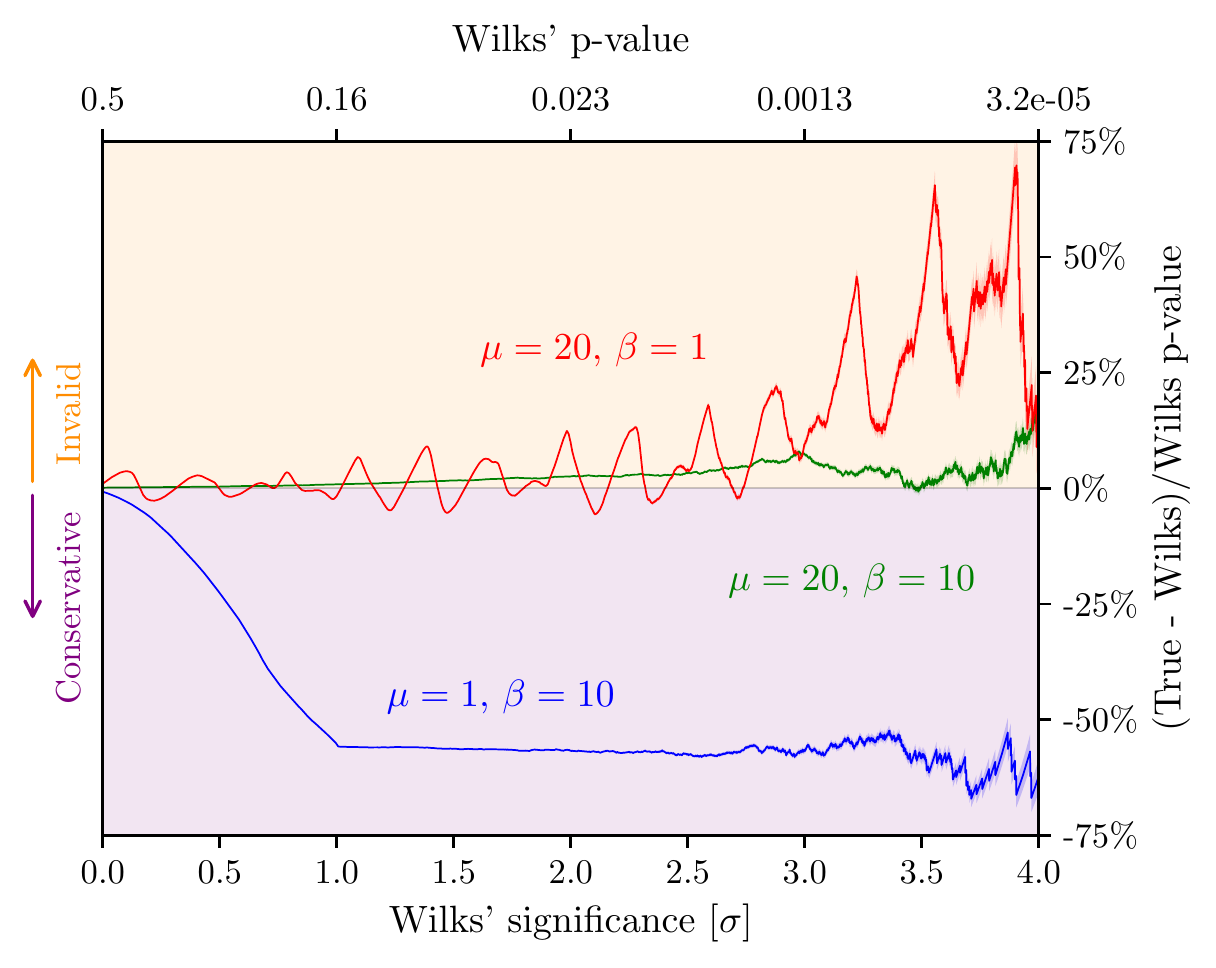}
\end{subfigure}%
\caption{Difference between true and Wilks-based p-values for the example experiment (see Figure~\ref{model_sketch} and eq.~\ref{L}) with a flat background ($B(E)=\phi_0$), no nuisance parameters, and  only allowing $\mu \geq 0$. \textbf{Left}: True p-value of $T$'s with a Wilks-based p-value of 0.1 ($T = \chi^2_{m=1}(90\%) \approx 2.71$), for different true $\mu$ and $\beta$. Each square on the figure represents the model at its center, based on $10^5$ toy simulations. \textbf{Right}: Relative error in Wilks-based p-values for different significance levels, based on $10^7$ toy simulations. Each curve corresponds to a model indicated with a dot of the same color on the left panel.}
\label{wilks_scan}
\end{figure*}
To illustrate these conditions, we consider an experiment that measured $N$ particles with energies $E^{(i)}$, with $i \in \{1,2,\dots,N\}$, and the interest is in the mean number $\mu$ of signal events on top of an expected background of $\beta$ expected events. The signal's energies are assumed to follow a Gaussian distribution with mean $\gamma$ and standard deviation 1, while the background distribution $B$ is polynomial, i.e.~$B(E) = k (\phi_0 + \phi_1 E + \phi_2 E^2 + \ldots)$, with $k$ a normalization constant, e.g., Figure \ref{model_sketch}.

For an (extended) unbinned analysis, the likelihood function is:
\begin{equation}
	\label{L}
L = \frac{1}{\mu + \beta} \text{Poisson}(N | \mu+\beta) \prod_{i=1}^N \Big[ \beta B(E^{(i)}) + \mu \text{Gauss}(E^{(i)} - \gamma) \Big] .
\end{equation}
Here, $\text{Poisson}(N|\mu+\beta)$ is the probability mass function of the Poisson distribution with mean $\mu+\beta$, and $\text{Gauss}$ the probability density function of a standard normal. 
For a discovery test, $H_0$ typically specifies as $\mu = 0$ and $H_1$ is either $\mu > 0$ (when only positive signals are allowed) or $\mu \neq 0$ (as in some neutrino oscillation experiments \cite{dayabay}). Finally, we consider as nuisance parameters collected in $\theta$ the set  $(\beta, \gamma, \phi_0, \phi_1, \ldots)$.

We can now describe each of the conditions necessary for  Wilks to hold.
\begin{itemize}
\item Technically, \wilkscond{asymptoticity} requires $N \rightarrow \infty$. We will consider practical requirements in the next section.
\item The true values of the both the parameters of interest and the nuisance parameters are in the \wilkscond{interior} of their respective parameter space.
For instance, under $H_0$, it would fail ~if only positive signals were allowed ($\mu \geq 0$).
\item
Each parameter  is \wilkscond{identifiable}--i.e., different values of the parameters specify different models. In our example, this holds if the signal location $\gamma$ is known. Whereas, the model is not identifiable when $\gamma$ is unknown and the signal is absent ($\mu = 0$).
\item In our example, $H_0:\mu = 0$ is a limiting case of $H_1:\mu>0$ (or $H_1:\mu\neq0$) hence we say that the models are \wc{nested}. This would not be the case, for instance, when testing $H_0 : \mu = 0$ versus $H_1 : \mu = 1$.
\item The experiment's model is \wc{correct}. It would fail e.g.~if the experiment has an additional unmodeled background component.
\end{itemize}

If the above-mentioned conditions hold, under $H_0$, the distribution of $T$ follows a $\chi^2_{m}$. In our example, $m=1$ since our hypotheses differ only in a single parameter $\mu$, i.e., the expected number of signal events.

\section{Failures of Wilks in practice}

\subsection{Insufficient data}
Physics searches, e.g.~for dark matter or neutrinoless double-beta decay, often look for rare signals on top of backgrounds they attempt to minimize. If the signal is weak or the background is low, e.g.~because the experiment ran only for a short time, there may be insufficient data for an \wc{asymptotic} approximation such as Wilks to be valid or to be able to discriminate the signal from the background distribution.

The left panel of Figure \ref{wilks_scan} shows the difference between true and Wilks'-based p-values for our example experiment, for a flat background $B=\phi_0$ and while increasing the number of signal events $\mu_0$ expected under $H_0$. The case of low background and moderate signal (red in Figure \ref{wilks_scan}), clearly illustrates the effect of a non-\wilkscond{asymptotic} test. The true significance for a certain $\chi^2$-threshold undulates around the desired significance, due to  the discrete number of observed events and ultimately the Poisson term in equation \ref{L}. Only if both $\beta$ and $\mu_0$ are large (green in Figure~\ref{wilks_scan}) Wilks' approximation becomes accurate.

\begin{figure}
      \includegraphics[width=85mm]{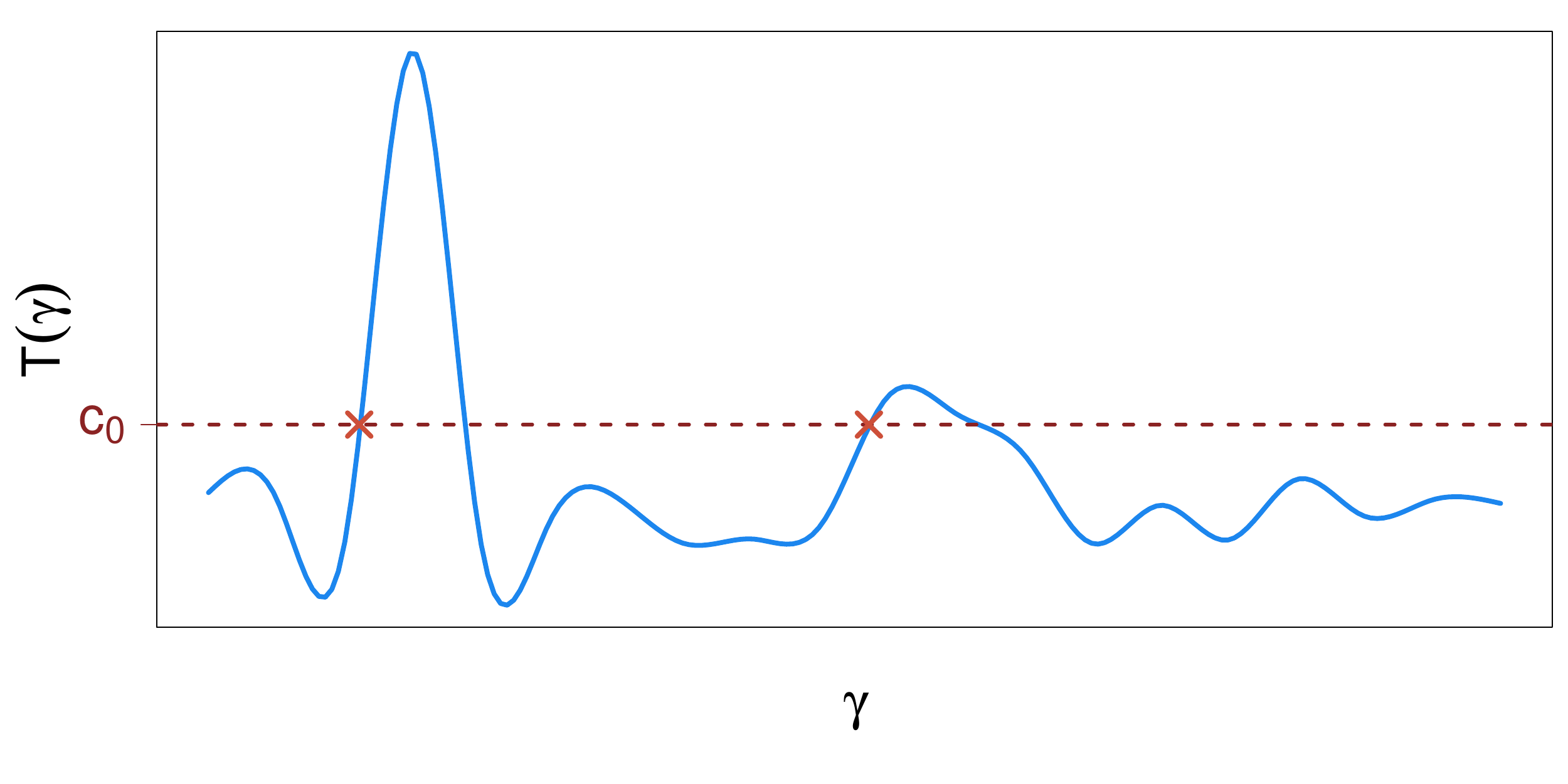}
      \caption{Upcrossings (red crosses) of a threshold $c_0$ by $\{T(\gamma)\}$ (blue line).}
      \label{upcr_fig}
\end{figure}
Mathematically, Wilks' requirement of asymptoticity ensures that the best-fit parameters, in our example $\widehat{\mu}$ and $\widehat{\theta}$, known as maximum-likelihood estimates (MLEs), are normally distributed with mean equal to their true value.
Consequently, the $\chi^2$ approximation derived by Wilks holds to order $O(N^{-1})$. Higher-order likelihood theory gives alternative test statistics whose distributions converge to $O(N^{-3/2})$, as reviewed comprehensively in \cite[Sec 2.2]{alessandra}. However, these statistics are often difficult to implement \cite{severini,he}; therefore, using simulations to estimate the null distribution of $T$ is often more practical.

%However, even experiments with much higher backgrounds may not be in an \wilkscond{asymptotic} regime, if the signal can be distinguished from the background to the point that the signal-like region becomes background-free.

% In the new-physics search example with no background ($\beta=0$), $\widehat{\mu} = N$, which is Poisson rather than normally distributed. The Poisson converges to a normal only for $\mu \rightarrow \infty$.

\subsection{Parameters with bounds}

\label{violation2}

%Experiments often constrain parameters with physical bounds, 
In many cases, parameters of statistical models may be constrained over a closed interval. Examples include particle masses or event rates that can only take positive values. In our example, this corresponds to the situation where $\mu \in [0,\infty)$. Therefore, $\mu=0$ is a boundary point  and thus, under $H_0$, $\mu$ is not in the \wilkscond{interior} of its parameter space. In this situation, Wilks theorem ultimately fails because the distribution of MLE is normal $50\%$ times  and it takes value zero on the remaining $50\%$.

%In our example, a model with high $\beta$ but low $\mu$ (blue in figure \ref{wilks_scan}) most clearly illustrates failure of \wilkscond{Interior}. In this case, $\widehat{\mu} = 0$ in about half the cases, leading to a low $T$ and high $p$. A sufficiently high $T$ (low $p$) is therefore about twice as rare as if there was no boundary, thus Wilks' p-values using Wilks' theorem are up to $50\%$ lower than the the true p-values (and therefore conservative).

%A boundary is not always obvious in the formulation of the problem. Consider our example with $l = \log \mu$ as the variable in the hypotheses. This transforms the problem of a boundary into one where the signal parameter is no longer \wilkscond{identifiable}.

%Although there is no explicit boundary, since $\mu = 0$ is at $l = - \infty$, the $T$ distribution is invariant to such changes of variable, and Wilks' theorem still fails.

When one or more of the parameters of interest are tested on the boundary of their parameter space, the limiting distribution of $T$ under $H_0$ is not $\chi^2$, but it may still enjoy a simple approximation. In our positive-signal example, the limiting $T$ distribution at the boundary $\mu=0$ is  $\frac{1}{2}\chi^2_{m=1}+\frac{1}{2}\delta(0)$, with $\delta(0)$ being the delta-Dirac  function centered at zero. 
This scenario was first studied by Chernoff\cite{chernoff} and later extended by others (e.g., \cite{selfliang}) to more general situations. 
Specifically, Self and Liang\cite{selfliang} show that, if the true values of the nuisance parameters also lie on boundaries, the distribution of $T$ under $H_0$ becomes more complex. In this setting, the limiting distribution may  differ over different regions of the parameter space and its asymptotic approximation may become particularly challenging (e.g., \cite[Case 7]{selfliang}).

Unfortunately, when the true values of the nuisance parameters  lie on the boundaries of the parameter space, simulations based on the MLE can lead to inconsistent results \cite{andrews2000}. However, estimators designed can be used instead of the MLE to avoid this problem \cite{andrews2000, charlie, cavaliere}.

%Finally, there exist situations
As a more specific example of a boundary, when the true value of $\mu$ moves close to the boundary in $\mu=0$ in figure~\ref{wilks_scan}, the MLE of $\mu$ ends up at the border more often, and the test statistic distribution morphs from Wilks' result to Chernoff's $\frac{1}{2}\chi^2_{m=1}+\frac{1}{2}\delta(0)$ at $\mu=0$. 
Toy Monte Carlo simulations may be necessary to determine if the signal is large/small enough to be at either extreme. This will also depend on the significance level desired, as shown for the model shown in blue in figure~\ref{wilks_scan} (with $\beta=10$ and $\mu_0=1$), where Chernoff's theorem approximately holds for significances corresponding to $p=0.1$, with $0.05$ of tests exceeding $\chi^2_1(90\%)$. This extends to the entire lower right region, with $\mu\lesssim1$ and $\beta\gtrsim5$ for confidence levels of $0.05$.

%$\widehat{\mu}$ is often zero\cite[p.142]{Book} and thus it does not follow a normal distribution.
% In our example, a model with $\beta=10$ and $\mu_0=1$ (blue in Figure \ref{wilks_scan}) most clearly illustrates this phenomenon. In this case, assuming that $\gamma$ is known, $\widehat{\mu} = 0$ in about half the cases, leading to low values of $T$ and high p-values. It follows that a sufficiently high value of $T$ (and consequently a low p-value) is therefore about twice as rare as if there was no boundary. In other words,  p-values obtained using Wilks' theorem are $50\%$ higher than the true p-values (and therefore more conservative).
%
\subsection{Non-identifiability and look-elsewhere effects}
Several theories of new physics often feature unknown parameters, such as the mass of a new particle.  For the model in \eqref{L}, this corresponds to the situation where the location of the signal, $\gamma$, is unknown. If the signal is absent ($H_0 : \mu = 0$), $\gamma$ is a \emph{non-identifiable} parameter: one that does not actually change the predictions of the experiment. Non-identifiability of a nuisance parameter implies that the limit of the respective MLE does not exist. Consequently, consistency and normality of the MLE are not guaranteed and  Wilks' theorem fails to hold. 

Searches for new  particles typically consider any value of $\gamma$ to which the experiment is sensitive. If the experiment wishes to report a separate confidence for each location, as may be the case with upper limits, a series of test statistics $\{T(\gamma)\}$, one for each value of  $\gamma$ fixed, can be employed. The most significant result (i.e.~the lowest p-value) can then be identified as an excess if it exceeds the field-specific thresholds.  

However, when several tests are conducted simultaneously the, \emph{look-elsewhere effect} (LEE) applies, i.e., the inference must be adjusted to preserve the desired (global) significance level $\alpha$, i.e., the probability of having at least one false detection among all the tests considered. Unfortunately, classical multiple hypothesis testing procedures such as Sidak's or Bonferroni's corrections (see \cite[Ch. 3]{efron}) either assume the tests are independent one another or are excessively conservative.
Unfortunately, the assumption of independence among the tests typically does not apply when $\gamma$ corresponds to the signal location as models testing for signals in nearby locations can only be weakly distinguished and thus the resulting tests would be highly correlated.

\begin{figure*}
\begin{adjustwidth}{0cm}{0cm}
\hspace{-0.5cm}
\begin{tabular*}{\textwidth}{@{\extracolsep{\fill}}@{}c@{}c@{}c@{}}
      \includegraphics[width=62mm]{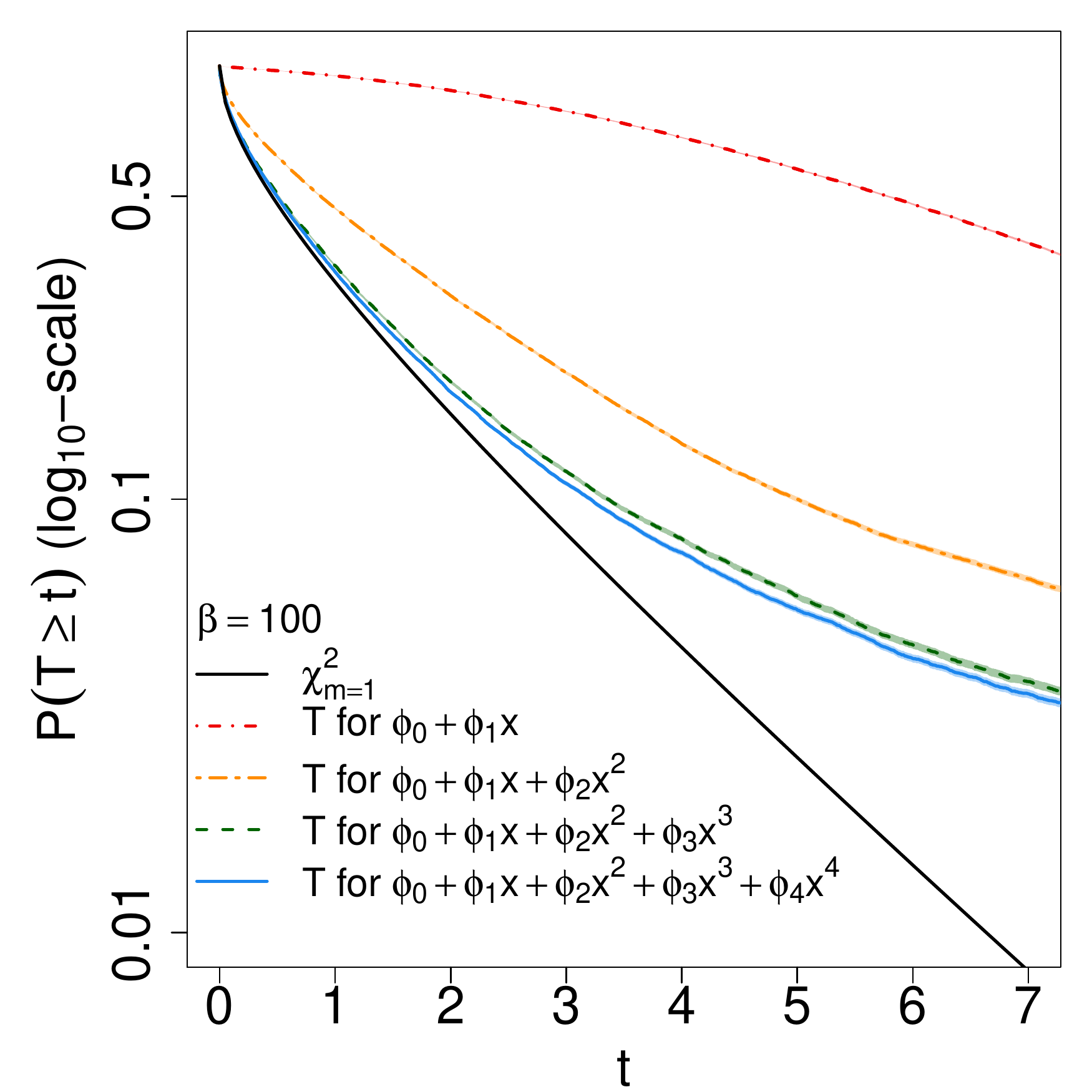} &\hspace{-0.2cm}\includegraphics[width=62mm]{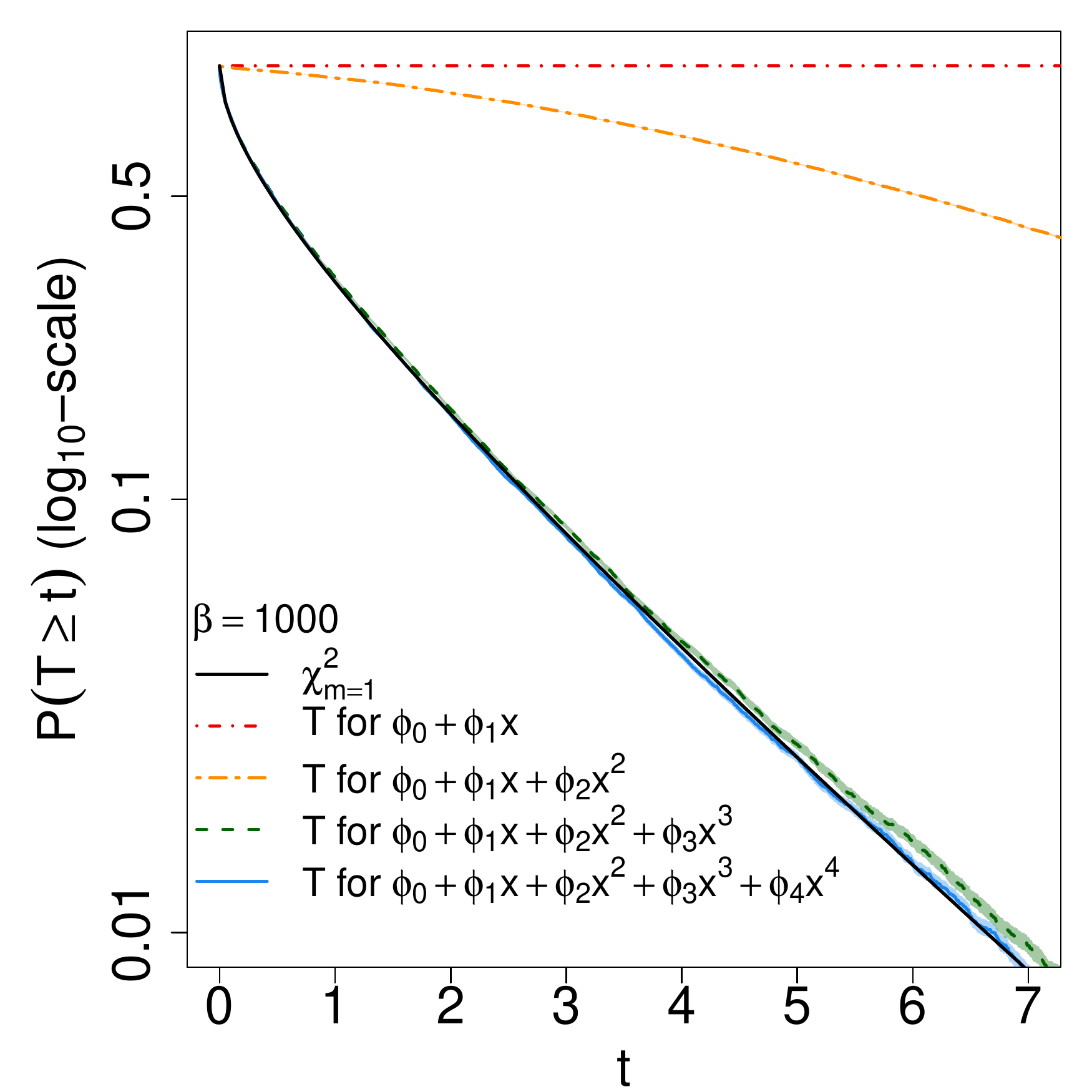} & \hspace{-0.3cm} \includegraphics[width=62mm]{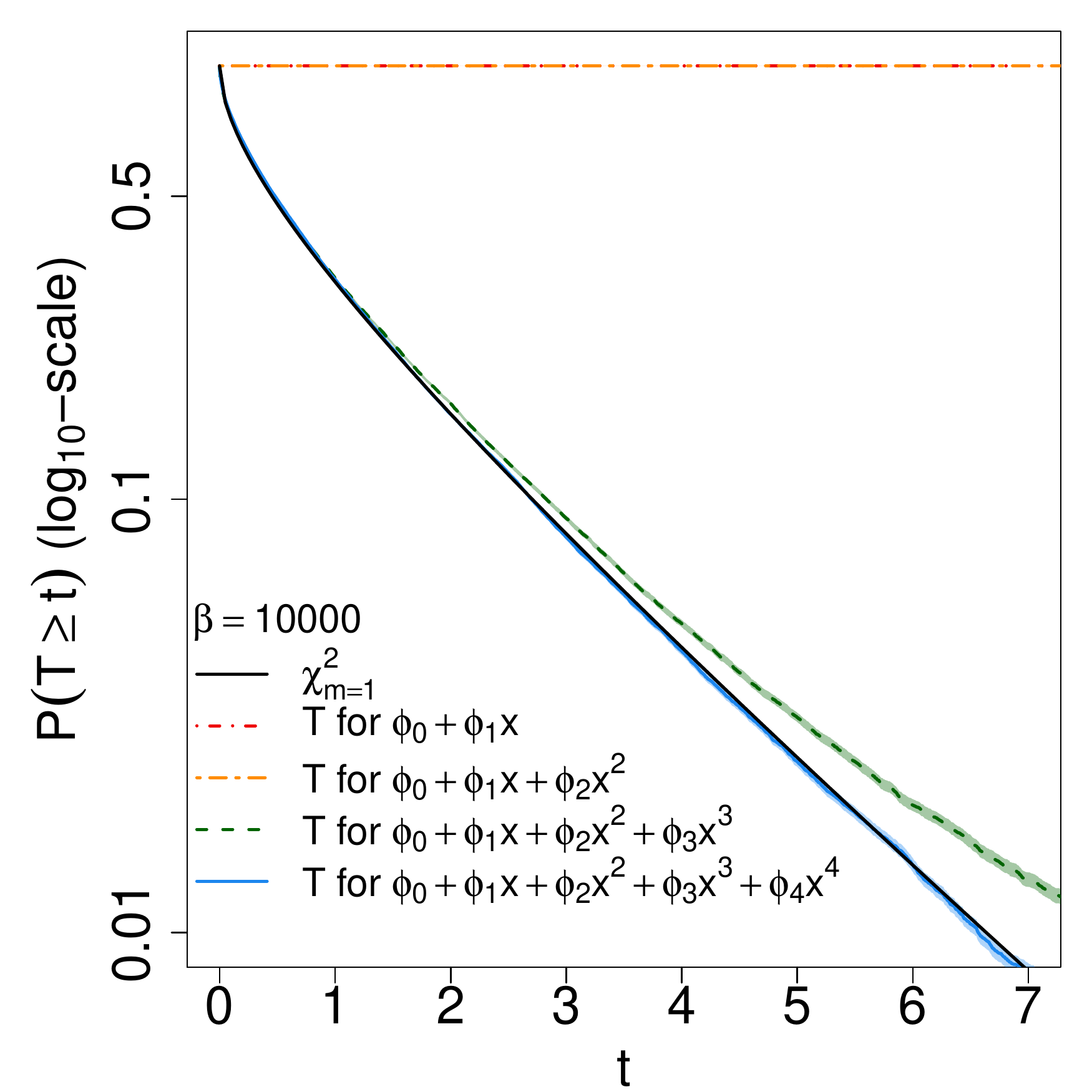} \\
\end{tabular*}
\end{adjustwidth}
\caption{
P-values approximations for \eqref{test} with $\mu_0=0$ when increasing the number of nuisance parameters and the expected number of events. The bias-variance trade-off plays a prominent role in approaching the  $\chi^2_{m=1}$ distribution (black solid line).
When $\beta=100$ (left panel) the sample size is not large enough to achieve a reliable $\chi^2_{m=1}$ approximation for none of the models considered. When increasing the expected number of events to $\beta=1000$ (central panel) both the $3^{rd}$ and $4^{th}$ degree polynomials (green dashed line and blue solid line, respectively) approach the $\chi^2$ approximation. In this case, despite the $4^{th}$ degree polynomial has  zero bias, it suffers from higher variance
than the cubic polynomial and thus they both lead to a reasonable $\chi^2$ approximation.
However, when $\beta=10000$ (right panel) only the $4^{th}$ polynomial leads to a reliable  $\chi^2_{m=1}$ approximation. }
\label{nuisance}
\end{figure*}

Gross and Vitells \cite{gv} introduced a novel approach to reduce the amount of Monte Carlo samples needed when nuisance parameters are not \wilkscond{identifiable} under the null hypothesis. Their method  aims to approximate the null distribution of  $max_\gamma\{T(\gamma)\}$ to derive a (global) p-value for the test $H_0:\mu=0$ versus $H_1:\mu>0$. 
When $\mu$ can take both positive and negative values, and assuming that the process $\{T(\gamma)\}$ is  distributed as a $\chi^2$ random process, the probability that $\max_\gamma\{T(\gamma)\}$  exceeds a high threshold $c$ can be approximated by the expected number of times $\{T(\gamma)\}$ upcrosses a lower threshold $c_0$ (see Figure \ref{upcr_fig}) and  denoted by $E[N_{c_0}|H_0]$. Specifically, as $c\rightarrow\infty$
\begin{equation}
	\label{upcross}
	P(\max_{\gamma}\{T(\gamma)\} > c | H_0) 
	\approx
	P(\chi^2_{m=1} > c) + E[N_{c_0}|H_0]e^{-\frac{(c-c_0)}{2}}      
\end{equation}
Despite $E[N_{c_0}|H_0]$ has to be estimated by simulation, it requires less Monte Carlo toys than those needed to estimate accurately the p-value of the high threshold $c$ directly. Furthermore, although the approximation in equation \eqref{upcross} is valid as $c\rightarrow\infty$, the right hand side bounds the left hand side from above for small values of $c$. Thus, the approximation yields to inference that might be overly conservative, but will not overstate an excess.

\subsection{Non-nestedness}
Many problems in physics involve the comparison of models which are not \wilkscond{nested}, i.e., one cannot be specified as a limiting case of the other. Examples include when testing the sign of an effect with known rate, or more commonly, when deciding between two incompatible functional forms such as (broken) power-laws and an exponential energy cutoff \cite{hesspks1745290}. The issue arising in this setting is that, despite the MLEs of the parameters of the two models are Gaussian, they do not share the same parameter space and thus $T$ is not guaranteed to be $\chi^2$ distributed.

A simple solution to the non-nested problem is to enlarge the model to one that covers both models under comparison as special cases.   For example, suppose the goal is to decide between two models $f$ and $g$ (e.g., two distributions of energies), each characterized by its own set of unknown parameters. We consider the comprehensive model $h = \eta f + (1 - \eta) g$. Here $0 \leq \eta \leq 1$ is a new parameter with no physical meaning but it is of crucial importance to perform inference. Specifically, two tests are constructed, i.e.,  
\begin{align}
    \label{t1}
\quad H_0:\eta=0\quad&\text{versus}\quad H_1:\eta>0\qquad\\
    \label{t2}
H_0:\eta=1\quad&\text{versus}\quad H_1:\eta<1.      \end{align}
and $f$ is selected if \eqref{t1} rejects $H_0$ but \eqref{t2} does not. Whereas, $g$ is selected if $H_0$ is rejected in \eqref{t2} but it is not in \eqref{t1}. In all the other cases the test is said to be inconclusive. 

In this construction $\eta$ is not in the \wilkscond{interior} of parameter space. Moreover, the parameters in $f$ and $g$ are not \wilkscond{identifiable} under $H_0$ in \eqref{t1} or \eqref{t2}. However, recent work \cite{algerinonnested,algeri2017,algeri2018,algeriDM} has shown that the result of Gross and Vitells can be extended to cover this case; the $\chi^2_{m=1}$ term in equation \eqref{upcross} acquires a factor $1/2$ because the parameter $\mu$ is tested on a boundary. Extensions of \cite{gv} and \cite{algerinonnested} to multiple dimensions are presented in \cite{vg,algeri2018,algeriDM}, whereas \cite{davies87,algeri2018} discuss the precise conditions that guarantee the validity of the resulting inference.

\begin{figure}
\centering
\includegraphics[width=\linewidth]{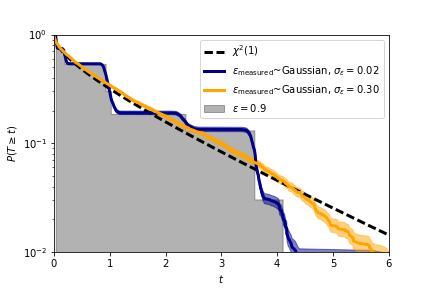}
\caption{Distribution of $T$ for a counting experiment with true signal expectation of $\mu = 2$ events, a known background of $\beta = 0.5$ events, and a detection efficiency $\epsilon = 0.9$. The efficiency is either considered known (gray), or as a nuisance parameter. In the latter case, a constraint term $\text{Gaussian}( (\epsilon - \epsilon_\text{measured})/\sigma_\epsilon )$ corresponding to the distribution of a subsidiary measurement $\epsilon_\text{measured}$ is added to the likelihood (eq.~\ref{L}). The $T$ distribution for $\sigma = 0.02$ (blue) and $\sigma = 0.3$ (orange) is plotted.} \label{fig:efficiency}
\end{figure}
 
It is important to point out that if $f$ and $g$ do not contain any unknown parameters, the problem is substantially simplified since, because of the Central Limit Theorem, the ratio of the likelihoods follows a Normal distribution\cite{cox62}.

\subsection{Uncertain models and nuisance parameters}

Experiments often have uncertainties on the nuisance parameters, such as detection efficiencies or background rates. As a result,  the models under either $H_0$ or $H_1$ may not be correctly specified and thus the last of our necessary conditions fails to hold.

A common remedy is to introduce additional nuisance parameters in order to increase the flexibility of the model and reduce its \emph{bias}, i.e., the discrepancy between the true underlying model and the model considered. On the other end, adding more parameters may substantially increase the \emph{variance} of the MLEs. Therefore, when specifying the background and/or signal models one must account for the trade-off between bias and variance.
 
To illustrate this phenomenon, we consider once again our example experiment. Suppose that the true (unknown) background is distributed as a fourth-order polynomial which we attempt to model with increasingly higher order polynomials. We aim to test $H_0:\mu=0$ versus $H_0:\mu\neq0$. In order to avoid boundary problems we allow $\mu$ to be negative. 

As Figure \ref{nuisance} shows,
when the experiment assumes a linear background (red chained lines), the  null distribution is substantially different from the one obtained assuming the true model (blue solid lines).
Whereas, as expected, the fit gets closer and closer to the truth when the polynomial order increases as this leads to a reduction of the bias. More interesting, however, is the effect played by the variance. 

Specifically, when the sample size decreases, the quadratic (orange chained lines) and cubic (green dashed lines) fits get closer to the one obtained with a fourth degree polynomial. This is because, the variance increases when the sample size decreases but it is reduced for models with a smaller number of parameters. This is further emphasizes by the fact that for small or only moderately large sample sizes, e.g., $\beta=100$ or $\beta=1000$ both the fourth and the third degree polynomial provide a similar fit.  However, when the expected number of events increases substantially ($\beta=1000$) the variance is dramatically reduced while the bias of the cubic fit is preserved. 

Remarkably, introducing nuisance parameters can sometimes help the achieve a $\chi^2$ approximation in non-asymptotic conditions, as illustrated in Figure~\ref{fig:efficiency}. Here, we consider an experiment with indistinguishable signal and background (a `counting experiment') and low expected counts. If the detection efficiency $\epsilon$ is known, $T$'s distribution is that of a (scaled) low-count Poisson, which deviates strongly from the asymptotic $\chi^2$-distribution. However, if $\epsilon$ is a nuisance parameter with a relatively broad constraint, the test statistic distribution is smeared out, and the asymptotic approximation performs better.

When the model imposes a functional form which is substantially different from the true model, e.g.~imposing a double exponential structure to fit data generated from a power law, additional nuisance parameters are unlikely to solve the problem. 
In this case, one can resort to nonparameteric methods\cite{algeri19}, or other conservative procedures \cite{safeguard, yellin} to correct mismodelling.

\newcommand{\tableok}{\textcolor{darkgreen}{\cmark}}
\newcommand{\tableno}{\textcolor{red}{\xmark}}
\newcommand{\tablemaybe}{\textcolor{orange}{\ding{58}}}
\begin{table*}
 \centering
\begin{tabular}{|c|c|c|c|c|c|}
 \hline
   &   &     &  &  & \\[-0.25cm]
 &{\bf\wc{ASYMPTOTIC}}&{\bf\wc{INTERIOR}}&{\bf  \wc{IDENTIFIABLE}}& {\bf \wc{NESTED}}  & {\bf \wc{CORRECT}} \\
   &   &     &  &  & \\[-0.25cm]
\hline
   && & & &  \\[-0.25cm]
Wilks' theorem  &\tableok&\tableok &\tableok & \tableok& \tableok \\

   && & & &  \\[-0.25cm]
\hline
   && & & & \\[-0.25cm]
Higher   order  asymptotics &\tableno&\tableok & \tableok&\tableok &\tableok  \\
   && & & &  \\[-0.25cm]
\hline
   && & & & \\[-0.25cm]
Boundary corrections   &\tableok&\tableno &\tableok &\tableok &\tableok  \\ 
    && & & & \\[-0.25cm]
  \hline
     && & & &  \\[-0.25cm]
LEE corrections &\tableok &\tablemaybe  & \tableno & \tableok & \tableok \\ 
   && & & &  \\[-0.25cm]
  \hline
     && & & &  \\[-0.25cm]
Tests for non-nested models  &\tableok&\tableok &\tableok &\tableno & \tableok \\ 

   && & & &  \\[-0.25cm]
  \hline
     && & & &  \\[-0.25cm]
Monte Carlo methods &\tableno& \tablemaybe& \tableno& \tableno&  \tableok\\ 
   && & & &  \\[-0.25cm]
               \hline
         && & & &  \\[-0.25cm]
Use of nuisance parameters       &\tablemaybe&
\tablemaybe&\tablemaybe &\tablemaybe&  \tablemaybe\\[-0.25cm]
           && & & &  \\
 \hline
      && & & &  \\[-0.25cm]
Nonparametric  methods      &\tableno&
\tableno&\tableno &\tableno &  \tableno\\[-0.25cm]
               && & & &  \\
            \hline
\end{tabular}
\caption{Inference under non-regular settings.  Green check marks indicate that the condition in the respective column must hold for the method on the left to be reliable. Orange plus marks indicate that extensions exists to cover situations where the respective condition does not hold (see text). Red cross marks indicate that the method on the left can be applied even if the  condition in the respective column does not hold.   }
\label{summary}
\end{table*}
\section{Recommendations}
\label{recommend}
When applying Wilks theorem, or more generally, when using the $\chi^2$ formalism   to calculate p-values or confidence intervals, one has to be aware of the regularity conditions that make this possible. Here we presented five conditions necessary for Wilks to hold  and which  should be sufficiently practical to serve as rules of thumbs on the applicability of the $\chi^2$ approximation for the LRT. Specifically, the conditions \wilkscond{interior}, \wilkscond{identifiable}, 
\wilkscond{nested} and \wilkscond{correct} refer to specific properties of the model under study. Thus, by studying the model in depth, it should be possible to understand whether they are fulfilled. The condition \wilkscond{asymptotic} refer to the size of the data available. Therefore, its validity is typically dictated by the specifics of the experiment being conducted. 

Table \ref{summary} outlines the solutions proposed in this paper to address the failure of each of the necessary conditions considered. Specifically, our recommendations can be summarized as follows.
\begin{itemize}
    \item If the data is not sufficiently large to guaranteed the validity of classical \wc{asymptotic} results, one can refer to approximations based on higher-order asymptotic likelihood theory or Monte Carlo simulations. However, the reader must keep in mind that the former  suffers from substantial mathematical complexity while the latter requires the availability of sufficient computational power.
    \item If the true values of the parameters are not in the \wc{interior} of their parameter space, one can implement boundary corrections for the p-values on the basis of the results of  Chernoff\cite{chernoff}, Self and Liang\cite{selfliang} and related works. Conversely, when aiming to address the problem by means of simulations, classical estimators of the nuisance parameters must be replaced by more efficient estimators\cite{andrews2000, charlie, cavaliere} to guarantee consistency of the solution.
    \item If nuisance parameters are not \wc{identifiable} and a fully simulated solution is excessively expensive, corrections for the LEE such as Gross and Vitells\cite{gv} and respective extensions allow to perform inference while reducing drastically the number of simulations required. 
    \item If the models under comparison are non-\wc{nested}, a simple solution is to specify a model which include the models under study as special cases\cite{algeri19}.
    \item If the likelihood specified is not \wc{correct}, one may attempt to recover the structure of the true underlying model by adding nuisance parameters and apply any of the above mentioned inferential methods. As any other modelling strategy, however, the bias-variance trade-off must be taken into account. 
\end{itemize}

Finally, if none of the assumptions above holds, or the correct models cannot be recovered by simply adding nuisance parameters, the only solution is to refer to nonparameteric methods (e.g.,\cite{algeri19}) for both statistical modelling and inference. 

\FloatBarrier

\bibliography{bibliography}

%\noindent \textbf{For Reviews only, highlighted references (optional)} \todo{Please select 5–-10 key references and provide a single sentence for each, highlighting the significance of the work.}
%
\section*{Acknowledgements }
J.C., J.A. and K.D.M. gratefully acknowledge support from the Knut and Alice Wallenberg Foundation, and the Swedish Research Council. 
%\todo{The authors thank Erwin the Cat for useful discussions. Please edit as necessary.}

\section*{Author contributions}
The authors contributed equally to all aspects of the article.

\section*{Competing interests}
The authors declare no competing interests.

\end{document}